\documentclass[twocolumn,amsmath,amssymb]{revtex4}

\usepackage{graphicx}
\usepackage{dcolumn}
\usepackage{bm}

\usepackage[ansinew]{inputenc}

\begin{document}

\title{InGaAs/GaAs/alkanethiolate radial superlattices: Experimental}

\author{Christoph Deneke}
\email{c.deneke@fkf.mpg.de}
\author{Ute Zschieschang}
\author{ Hagen Klauk}
\author{Oliver G. Schmidt}
\affiliation{Max-Planck-Institut f\"ur Festk\"orperforschung, Heisenbergstr.1, D-70569 Stuttgart, Germany}

\date{\today}

\begin{abstract}

A radial InGaAs/GaAs/1-hexadecanethiol superlattice is fabricated by the roll-up of a strained InGaAs/GaAs bilayer passivated with 
a molecular self-assembled monolayer. Our technique allows the formation of multi-period inorganic/organic hybrid 
heterostructures.  This paper contains the detailed experimental description of how to fabricate these structures.

\end{abstract}

\maketitle

We describe how to produce semiconductor/organic superlattices by rolling up a strained semiconductor bilayer functionalized with 
a molecular self-assembled monolayer (SAM). The roll-up procedure enables us to compose alternating layers of single-crystalline 
inorganic semiconductor heterostructures (InGaAs/GaAs) and functional organic layers (thiolate SAM) in a radial geometry. 
Cross-sections of such RSLs are carefully prepared by targeted focused ion beam (FIB) etching and polishing~\cite{deneke2006}. 
This refined technique enables us to study the structure and geometry of well-selected RSLs in unprecedented detail. 
High-resolution and analytical transmission electron microscopy ((HR)TEM) reveal tightly wound RSLs with high-quality interfaces 
and precisely controllable layer thicknesses and chemical compositions~\cite{apl}.

\section{RSL preparation}
To obtain inherently strained bilayers, 2 nm In$_{0.33}$Ga$_{0.67}$As and 4.5 nm GaAs were grown on top of a 19 nm thick AlAs 
layer by molecular beam epitaxy (MBE) on a GaAs (001) surface. After growth the sample was removed from the MBE machine and a 
self-assembled monolayer (SAM) of 1-hexadecanethiol (HDT) was allowed to self-assemble on top of the GaAs. Prior to molecular 
self-assembly the sample was etched for 30 sec in concentrated HCl, rinsed with DI water and methanol and immediately immersed in 
a methanol solution containing 5 mM HDT. The HCl etching step removes the native oxide and parts of the top GaAs layer resulting 
in a thinning of the top GaAs layer by a few \r{A}ngstroms. After 24h at room temperature the sample was removed from the HDT 
solution, rinsed with methanol and dried in a stream of nitrogen. To release the InGaAs/GaAs/thiolate trilayer from the GaAs 
substrate, the sample was scratched and etched for 20 to 30 sec in an HF (50 vol\%):H$_2$O (1:15) solution removing the AlAs 
sacrificial layer. The reference sample was processed in the same manner except that the InGaAs/GaAs bilayer was released 
immediately after the HCl etching step.
The formation of a high-quality SAM was verified by measuring the contact angle of water at the surface before (32$^{\circ}$) and 
after (96$^{\circ}$) SAM preparation. We emphasize that the contact angle was stable against the exposure of a reference sample to 
concentrated HF.

\section{FIB preparation}
To prepare cross-sections of the rolled-up nanotubes focused ion beam (FIB) techniques were used. First, the region of interest 
was covered by a metal stripe (W or Pt). Next, a lamella was prepared out of the sample at the desired position using a focused Ga 
beam. The lamella was attached to a TEM grid and thinned to electron transparency. During thinning of the sample great care was 
taken to prevent damaging and amorphisation of the sample. To ensure electron transparency suitable for high-resolution imaging, 
one tube was inhomogeneously thinned until a section of the tube was completely removed.

\section{TEM investigation and chemical analysis}
The (S)TEM images and chemical analysis of the InGaAs/GaAs/thiolate radial superlattices was recorded in a FEI Tecnai G2 TF20XT 
operated at 200 kV. The instrument was equipped with a GIF Tridiem Gatan imaging filter, an EDAX EDS spectrometer and a 
high-angular darkfield (HAADF) detector for STEM imaging. The chemical analysis was performed in nanoprobe STEM mode by 
simultaneous recording of EDX, EELS and HAADF signals. The reference InGaAs/GaAs sample was investigated with an energy-filtered 
Zeiss LIBRA200 FE TEM at 200 kV. Microdiffraction was carried out at a selected area of the rolled-up InGaAs/GaAs layer stack 
during the TEM investigation

\section{Conclusion}

We have created multi-period semiconductor/organic radial superlattices by the roll-up of highly strained 
InGaAs/GaAs/alkanethiolate trilayers. Detailed cross-sectional TEM and chemical analysis shows that the radial superlattices 
consist of alternating crystalline semiconductor and organic layers with high-quality interfaces that are practically free of any 
contaminations (see Ref.~\onlinecite{apl}). Since a strained crystalline layer can be combined deliberately with arbitrary 
combinations of organic or inorganic materials, our radial superlattices define a class of hybrid short-period heterostructures, 
which can be integrated on a single chip by standard lithographic techniques.

\end{document}